\documentclass[]{emulateapj}
\usepackage{graphicx}
\shorttitle{Polarization intermittency}
\begin{document}
\title{Polarization intermittency
in MHD and its influence
on turbulent cascade}            
\author{A. Beresnyak, A. Lazarian}
\affil{Dept. of Astronomy, Univ. of Wisconsin, Madison, WI 53706}
\email{andrey, lazarian@astro.wisc.edu}

\begin{abstract} 
Goldreich-Sridhar model of incompressible turbulence provides
an elegant approach to describing strong MHD turbulence. 
It relies on the fact that interacting Alfv\'enic
waves are independent and have random polarization. However, in case
of strong interaction a spontaneous local assymetry can arise.
We used direct numerical simulations to show that polarization
alignment occurs and it grows larger at smaller scales.
This effect would lead to a shallower spectrum and stronger
anisotropy. Even small changes in these two properties will have
important astrophysical consequencies, e.g. for the cosmic ray physics. 
\end{abstract}


\section{Introduction}
Astrophysical plasmas is turbulent and magnetized. Magnetohydrodynamic
turbulence affects many phenomena including the propagation and acceleration
of cosmic rays (see Shlickeiser 2003). While foundations of the theory
of anisotropic MHD turbulence model can be traced back to works in 80s
(see Shebalin, Matthaeus \& Montgomery 1983, Higdon 1984, Matthaeus \& Brown 1988) a 
substantial recent progress was initiated by the pioneering study by
Goldreich \& Sridhar (1995, henceforth GS95). There a concept of balancing of linear
and non-linear term in incompressible MHD equations was suggested, which
results in a prediction between the wavenumbers parallel $\|$
and perpendicular $\bot$ to the local direction of
magnetic field $k_{\|}\sim k_{\perp}^{2/3}$. GS95
model also predicts Kolmogorov $E(k)\sim k_{\bot}^{-5/3}$ 
spectrum, which is consistent with both
interstellar (see Armstrong, Rickett \& Spangler 1995) and Solar wind (see 
Horbury 1999) data.

While the core concept of the GS95 model, namely, critical balance,
was confirmed in numerical simulations (Cho \& Vishniac 2000, 
Maron \& Goldreich 2001, Cho, Lazarian \& Vishniac 2002, henceforth  CLV02), 
the simulations revealed a 
difference in the spectral indexes obtained. For instance, Maron \& Goldreich
(2001) obtained the spectral index that is close to $-3/2$ in
contrast to $-5/3$ in CLV02. Index of $-3/2$ for field-perpendicular energy
was obtained in simulations of Muller \& Grappin (2005). This stimulated
theoretical efforts
to understand the difference between the theory and simulations on one hand
and between different sets of simulations on the other hand. In particular,
Boldyrev (2005) proposed a model in which the interactions are weakened 
compared to the GS95 predictions as a result of 3D structure of eddies.
An alternative point of view, namely, flattening of the spectrum as the
result of intermittency was discussed in Maron \& Goldreich (2001).

We note, that the problem of MHD turbulence spectrum is of great practical
importance. For instance, the introduction of GS95 spectrum and its
extension to compressible media (see Lithwick \& Goldreich 2001,
Cho \& Lazarian 2002, 2003) resulted in a substnatial changes in understanding
of cosmic ray propagation and acceleration (see Chandran 2000, Yan \& Lazarian
2002, 2004, Cho \& Lazarian 2006). 

In what follows we address the issue of interaction efficiency using numerical
simulations. We use both compressible and incompressible MHD runs. Contrary
to some earlier claims, Cho \& Lazarian (2002) demonstrated weak coupling
of the compressible and Alfvenic parts of the spectra, which justifies
comparing Alfvenic turbulence in incompressible and compressible media.

\section{Incompressible MHD}
In the incompressible case it is convenient to use Els\"asser variables
${\bf w}={\bf v}_A+{\bf v}-{\bf b}$, ${\bf u}=-{\bf v}_A+{\bf v}+{\bf b}$,
where ${\bf v}_A$ and ${\bf b}$ are mean and total magnetic fields
in velocity units (see Biskamp 2003). With those MHD equations will have
a symmetric form

$$
\partial_t{\bf u}-({\bf v}_A\cdot\nabla){\bf u}=-({\bf w}\cdot\nabla){\bf u}-\nabla P, \eqno (1)
$$

$$
\partial_t{\bf w}+({\bf v}_A\cdot\nabla){\bf w}=-({\bf u}\cdot\nabla){\bf w}-\nabla P, \eqno (2)
$$

$$
\nabla\cdot{\bf u}=\nabla\cdot{\bf w}=0. \eqno (3)
$$

Here the total pressure $P=p+B^2/8\pi$ is determined by incompressibility
conditions (3). Explicitly

$$
P=\int\frac{d^3x'}{4\pi}\frac{\nabla{\bf w}:\nabla{\bf u}}{|{\bf x'}-{\bf x}|}.
$$

Aside from symmetry, the remarkable property of these equations is
the existence of the exact nonlinear solutions for one field,
such as ${\bf u}={\bf f}({\bf r}+{\bf v}_A t)$, in the absence of
the other, ${\bf w}={\bf 0}$.

Variables ${\bf u}$ and ${\bf w}$ can be Fourier decomposed into
waves which have a dispersion relation of $\omega=v_A k_\|$. These
are actually consist of two modes, shear Alfv\'en waves, which have
${\bf u}$ and ${\bf w}$ perpendicular to both ${\bf k}$ and
${\bf v}_A$, and pseudo-Alfv\'en, or slow waves that in incompressible
fluid compress magnetic field. It have been shown both
euristically and numerically, that in developed turbulence shear
waves govern the cascade and the back-reaction of slow waves are
relatively unimportant (see Maron \& Goldreich 2001). 

In equations (1) and (2), written in Fourier space, linear term
could be estimated as $v_A k_\| u$ and nonlinear as $w k_\perp u$.
With sufficiently small wave amplitudes linear term dominates
and the theory of so called weak or wave turbulence could be built
(Galtier et al, 2002). Due to the peculiar nature of the dispersion
relation only waves with opposite and equal by magnitude 
$\bf{k}_\|$ can interact, to conserve momentum. Thus, only
perpendicular cascade ensues, which leads to the waves with
sufficiently large $k_{\perp}$, so that nonlinear term is no
longer small. It was proposed in GS95 that from this point
turbulent cascade evolves keeping approximate balance between
linear and nonlinear terms, so called critical balance.
The waves with larger $k_{\|}$ are created by decorrelation in the lateral
structure ot the wave packet because of the strong interaction.

In the weak turbulence it is safely to assume that ${\bf u}$ and
${\bf w}$ wave packets have independent polarization, since they
interact weakly, only once, and never meet again. In strong
turbulence this is not nessesarily true as the head of the 
${\bf u}$ wavepacket got significatly modified as it reaches the tail
of the ${\bf w}$ wave packet. Futhermore, as we speak of the Alfv\'en
mode, the nonlinear term is actually proportional to 
$w k_{\perp} u \sin\theta$ where $\theta$ is an angle between
${\bf w}$ and ${\bf u}$. Therefore those wave packets that have
nearly parallel polarizations can survive for longer, their cascading
being inhibited. 

\section{Numerical codes}
We used several sets of turbulent data flows, produced in driven direct
MHD numerical simulations. We used both incompressible pseudospectral
code, which is described in detail in CLV02, and
compressible isothermal ENO code, described in Cho \& Lazarian 2002.
In both cases we used periodic $512^3$ grid, isotropic random divergentless
driving with correlation time around unity in Alfv\'en crossing times.
Three incompressible simulations had Alfv\'enic Mach number, calculated
as the ratio of the kinetic to magnetic energy squared, $M_A=0.7$, $1.0$
and $1.4$. Compressible run had $M_A=1.0$ and sonic Mach number of around
unity.

\section{Results}

The sample spectra of the magnetic and kinetic energies
are given in Fig. 1.

%\placefigure{spectrum}
\begin{figure*}
\figurenum{1}
%  \plottwo{sp.ps}{sp1.ps}
  \includegraphics[width=0.49\textwidth]{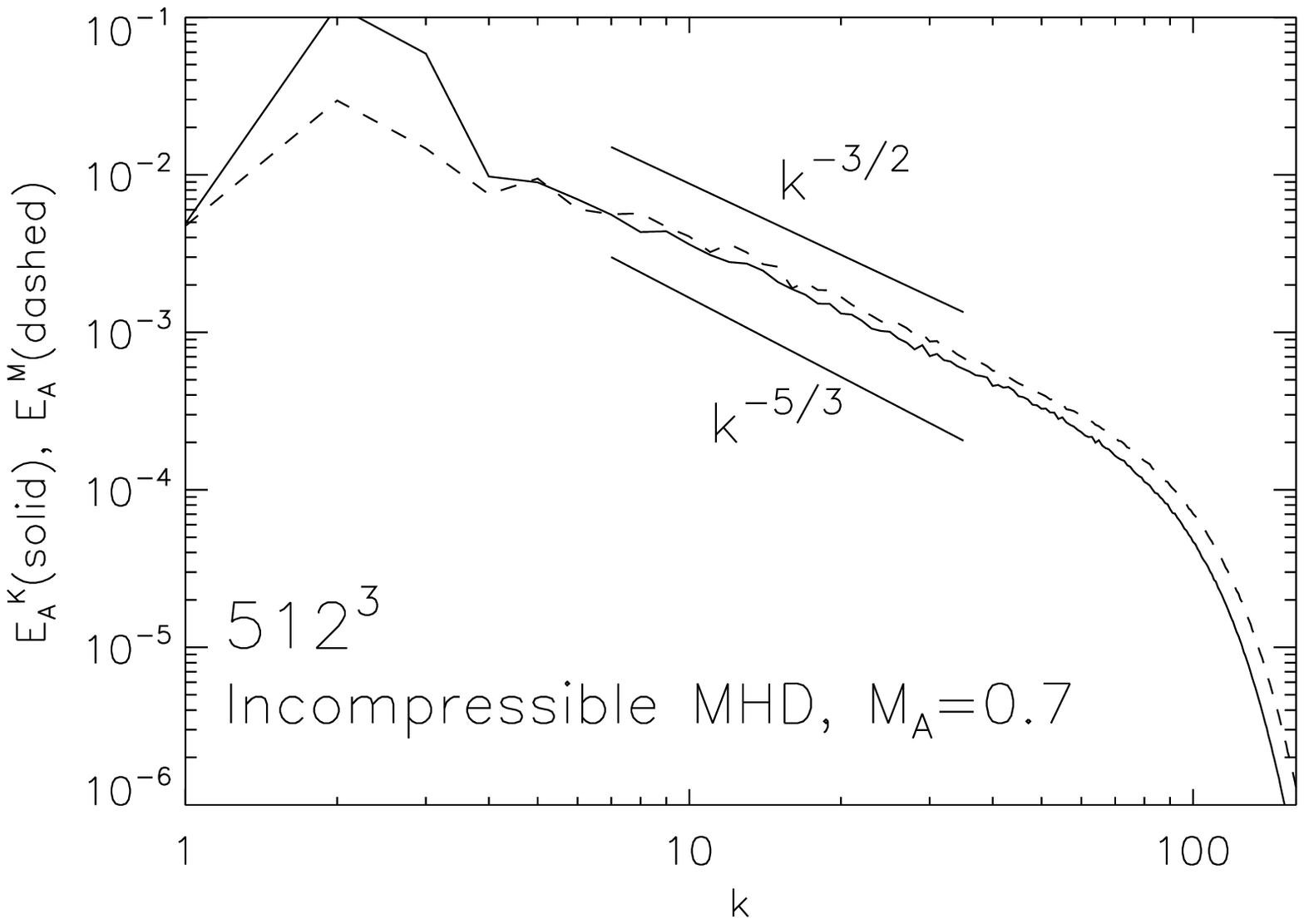}
  \hfill
  \includegraphics[width=0.49\textwidth]{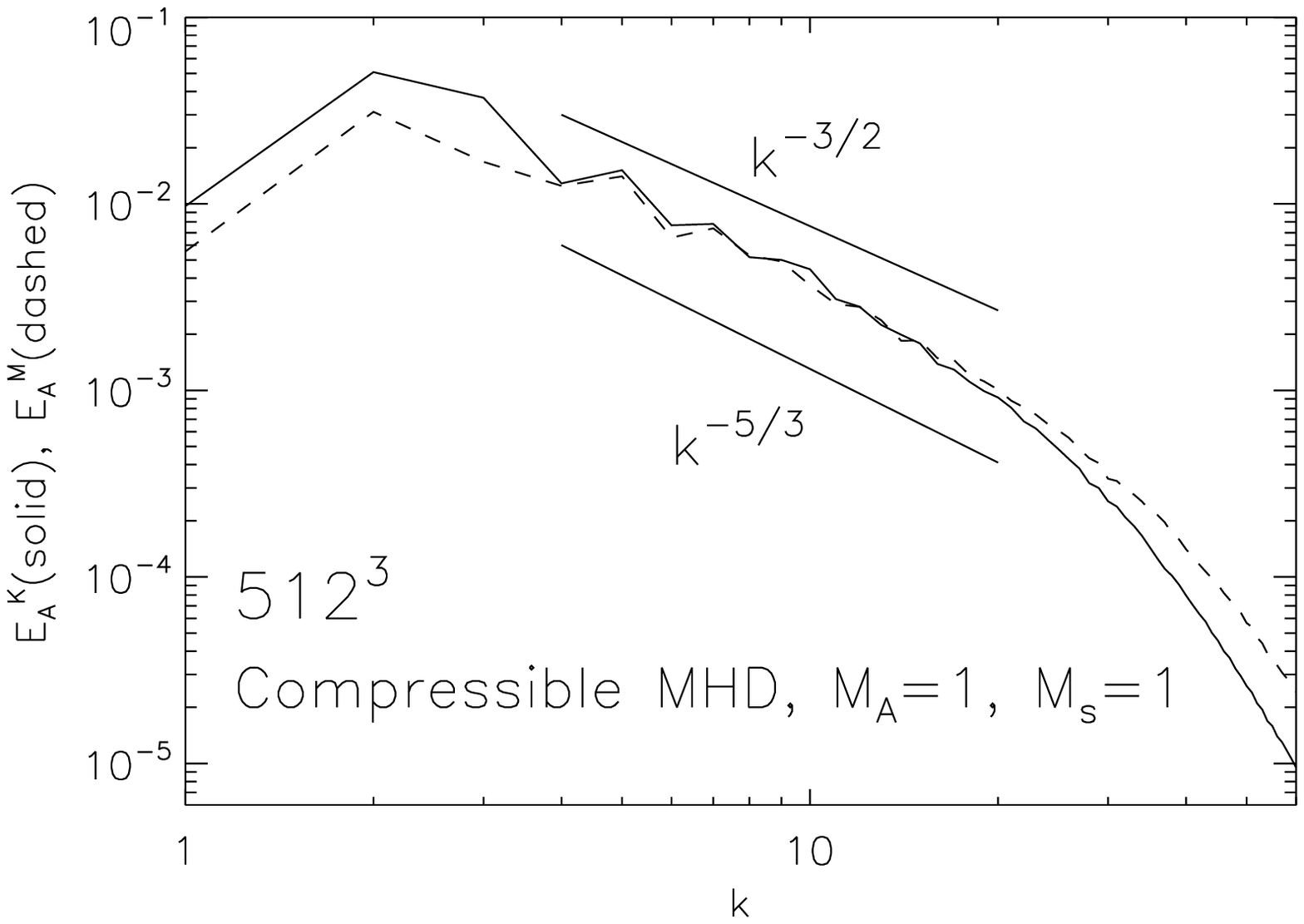}
  \caption{Spectra of kinetic (solid) and magnetic (dashed) energies
 of Alfvenic mode. {\it Left}: incompressible MHD, $M_A=0.7$, hyperdiffusion
of the third order, $\nu_3\Delta^3$,
{\it right}: compressible run with $M_A=1$, $M_s=1$.}
\label{spectrum}
\end{figure*}

We are asking the question of whether polarizations
of Alfv\'enic parts of $w$ and $z$ are truly independent,
or they have significant correlation. Since the effect we are
looking for could be small, we used two different methods
of separation of scales and modes in order to assure that
the effect is not due to spurious correlation which came
e.g. from uncertainty in mode decomposition.

The first method was to obtain datacubes filtered in 
Fourier space, where only wavevectors around some $k_{\perp 0}$
and $k_{\| 0}$ were left. The uncertainty in $k_{\perp}$ was
of the order of $k_{\perp 0}$, while $k_{\| 0}$ and its uncertainty
was set according to GS95 anisotropy.
We took the Alfv\'enic mode which has vectors
perpendicular to both ${\bf k}$ and {\it mean} ${\bf B}$.
After filtering we tranformed fields into real space. We checked
that their mean square values were of the order of unity,
as it has to be for the strong and local turbulence.

We than proceeded to calculate an rms value of the
estimated nonlinear term $w k_{\perp} u |\sin\theta|$,
the rms of the angular anisotropy, which is $<\sin^2\theta>^{1/2}$,
both compensated by $2^{1/2}$,
and the estimated nonlinear term without $\sin\theta$, or $w k_{\perp} u$.
We refer to the ratio of the first and the third as the weakening
of the interaction corresponding to wavevector $k$, as this is the number
in which nonlinear term is smaller than a naive estimate based
on independency of polarization for ${\bf w}$ and ${\bf u}$.
We refer to the second term as the measure of geometrical
alignment of polarization. It will be unity if $w$ and $u$ are not
aligned.

The second method was one of a transverse structure functions (TSF),
calculated with respect to the {\it local} magnetic field. Namely,
we calculated the square root of the forth order structure
function $2<|\delta w \delta u \sin\theta|^2>$, where the $\delta$
difference is taken between two points, connected by vector ${\bf l}$ 
perpendicular to the local magnetic field, and $\theta$ is an angle between
$\delta {\bf w}$ and $\delta {\bf u}$, and similar
structure functions $2<|\sin\theta|^2>$,
and $<|\delta w \delta u|^2>$. And again we call the square root
of the second TSF the geometrical alighnment and the ratio of square roots
of the first and the third as the weakening of interaction, corresponding
to scale $l$. Here we replaced full $w$ and $z$ with their projections
on the plane perpendicular to the ${\bf B}$ vector.

The above two methods have different uncertainties that go with them.
The first method uses {\it global} mean magnetic field to do both
mode decomposition and separation of scales. The second method uses
{\it local} magnetic field for separation of scales, as we use ${\bf l}$
perpendicular to the local field, but it uses {\it approximate} method for
{\it mode separation}, as there is some admix of the slow mode in the vectors
perpendicular to B. This admix, however, has to be small on small scales
as the wavevector tends to be mostly perpendicular to the magnetic field
hence the vector of the slow mode is mostly along magnetic field. 

If the effects of alighnment are not spurious we will see correspondence
of two methods when $k \sim 1/l$. We have plotted the geometrical
alighnment factor and the weakening of the interation factor for
some of our datacubes on Fig. 2. As we see, there is a
correspondence between two methods of calculating our factors.

%\placefigure{SF}
\begin{figure*}
\figurenum{2}
  \includegraphics[width=0.32\textwidth]{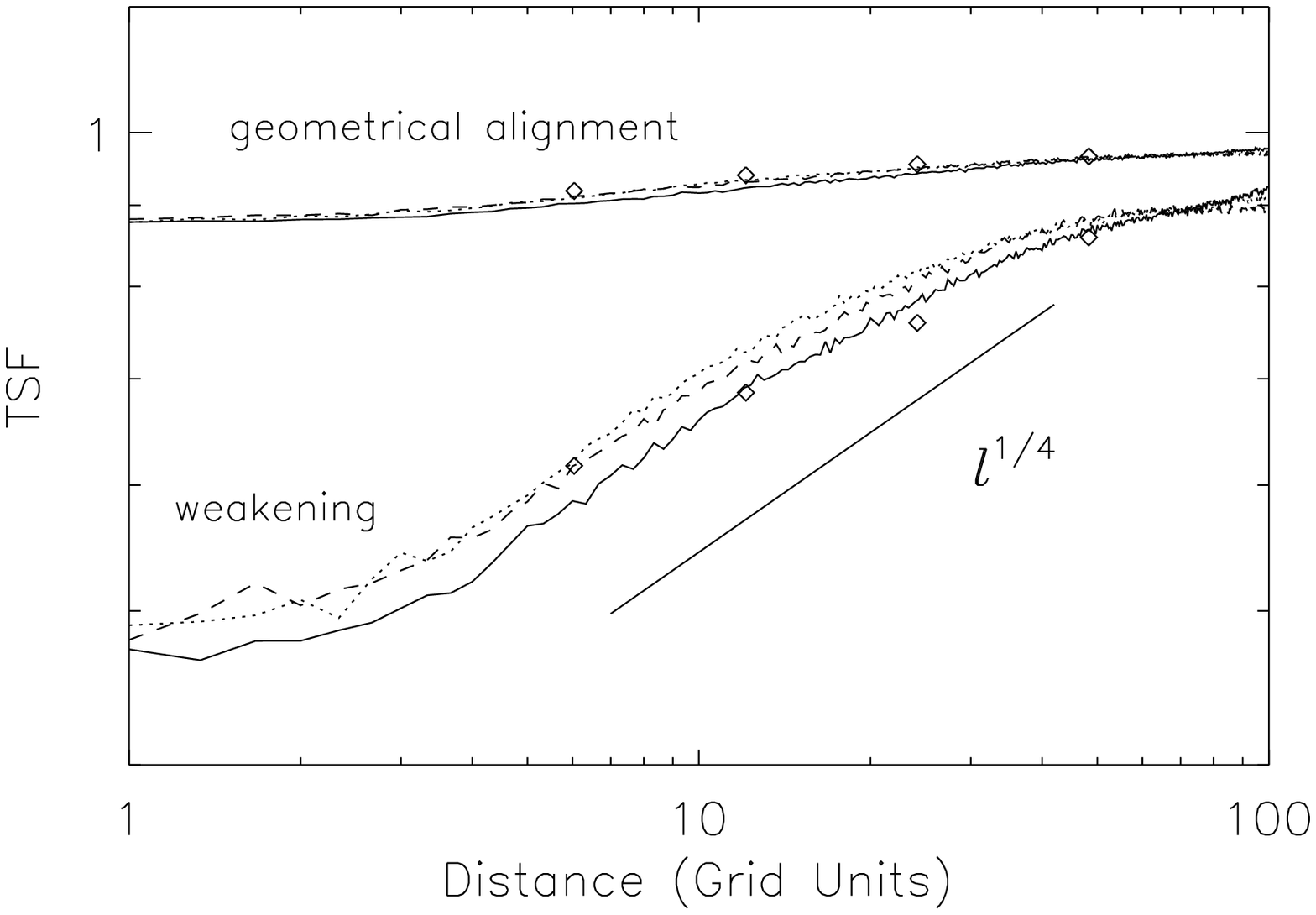}
  \hfill
  \includegraphics[width=0.32\textwidth]{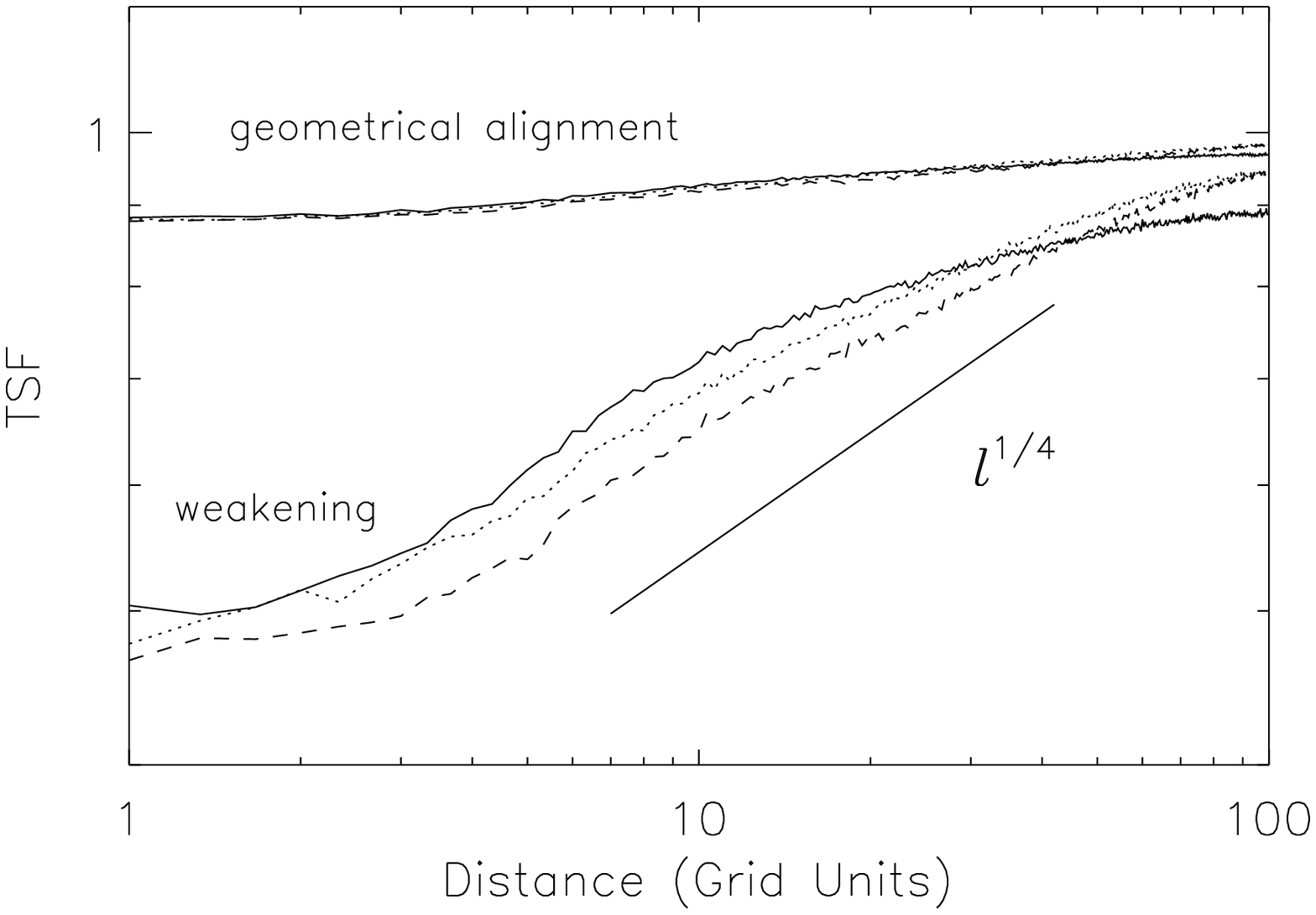}
  \hfill
  \includegraphics[width=0.32\textwidth]{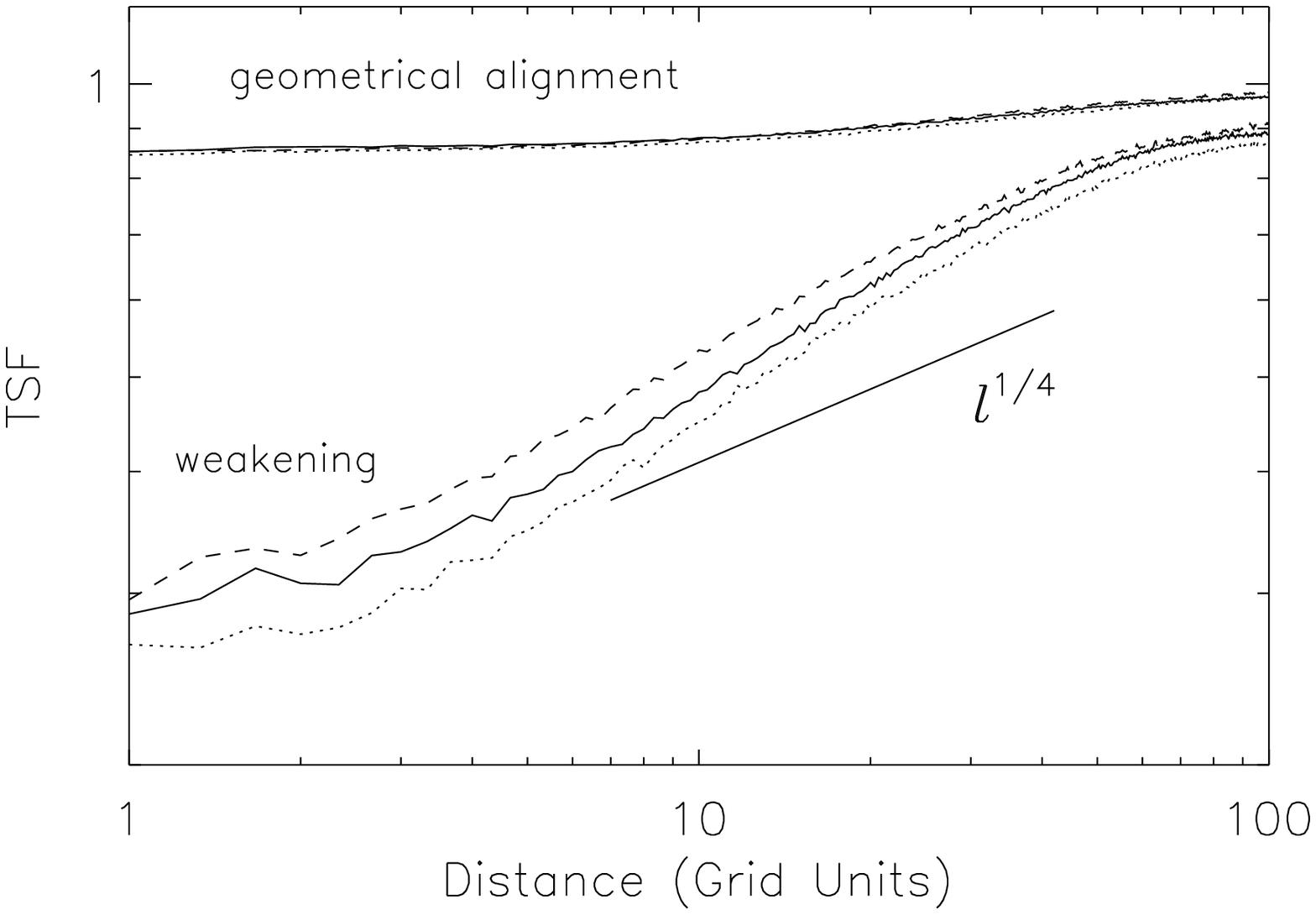}
  \caption{ The geometrical alignment (upper curve) and the weakening of
 interaction factor (lower curve), {\it left panel}: incompressible, $M_A=0.7$,
comparison of two methods refered in text, diamonds are for the first method,
they were put on distance scale $l \sim 0.7/k$, lines are for second
method (TSF). Solid, dotted and dashed lines are for three simulations,
separated by Alfv\'enic cross time. {\it Central panel}: incompressible, $M_A=1.0$,
same notation. {\it Right panel}: Compressible, $M_A=1$, $M_s=1$,
we used a global mean magnetic field for both mode separation and scale separation.}
\label{SF}
\end{figure*}

We see that a purely geometric alighnment
factor $2^{1/2}<\sin^2\theta>^{1/2}$ is quite close to unity. It is certainly
not enough to produce significant interaction
weakening. On the other hand, the ``weakening factor'' is sizable, being
aroung 0.5 for smallest scales. This means that there is a significant
correlation between wave amplitudes and the polarization alignment.
In other words, polarization alignment is rather strong where there is
a large turbulent field. We additionaly confirmed
this by plotting the distribution of angles between Alfv\'en ${\bf w}$ and
${\bf u}$ for regions with different wave amplitudes. It seems that
in high-amplitude
regions the alignment is much stronger. Therefore we see that the
effect has an intermittent nature.

\section{The Model}
The intermittent weakening of the interaction can in principle
show itself in different ways. We follow a particular model that
assume, that critical balance is persistent. Indeed, if, in one
region, interaction become weakened it is compensated by further
growth of $k_{\perp}$ while $k_{\|}$ does not grow much because
there is no significat lateral decorrelation. 

Let's assume that weakening factor scale as $k_{\perp}^{-\alpha}$.
Then, a critical balance would mean that 
$v_A\delta u k_{\|} \sim \delta u^2 k_{\perp}^{1-\alpha}$. Here we use
$u$ instead of both $u$ and $w$, as we consider balanced case.
Then cascading time could be determined by either linear or
nonlinear term as $\tau \sim \delta u^{-1} k_{\perp}^{-1+\alpha}$.
The Kolmogorov hypothethis $\delta u^2/\tau=const$ will give
us the spectrum of $\delta u\sim k_{\perp}^{(-1+\alpha)/3}$ or
$E_k \sim k_{\perp}^{-5/3+2\alpha/3}$. The anisotropy is obtained
from critical balance: $k_{\|} \sim k_{\perp}^{2/3-2\alpha/3}$.

In case of $\alpha=0$, scale independent weakening factor, we
reproduce GS95 spectrum and anisotropy. In order to achieve
Iroshnikov-Kraichnan spectrum we have to take $\alpha=1/4$.
However in this case anisotropy will be stronger than in GS95,
namely $k_{\|} \sim k_{\perp}^{1/2}$.

\section{Intermittency}
In section 2 we chose to use 4th order structure functions (SF) to
quantify interaction weakening. This choice is somewhat arbitrary.
Indeed, the Kolmogorov-type arguments suggest using 3rd order
functions (see Monin \& Yaglom 1975), but, alas, we cannot naturally
construct such a structure
function in MHD (for discussion, see Biskamp 2003, sec. 7.3.3).
If we assume that weakening is proportional to the wave amplitude,
it is more natural to use 4th order SF. We desided to calculate the relative
scaling exponents for combined Els\"asser fields, as well as nonlinear
term. This will quantify how much depletion of interaction or
``interaction weakening'' that we calculated, depends on the choice
of the order of the SF.

One of the curious things we have found is that extended self-similarity
(ESS, see Benzi et al., 1993) between SFs of different order is
much better for nonlinear term than for combined elsasser fields.
For the latter, self-similarity between SFs are mostly limited to
inertial range. Obviosly there is no good extended similarity between
nonlinear term and elsasser fields. We shall cautiosly conclude here,
that nonlinear term, $wu\sin\theta$ is more ``fundamental'' than $wu$.

We have plotted relative scaling exponents for nonlinear term
and the product of Els\"asser fields on figure 3. We used 4th order
SF $w^2u^2\sin^2\theta$ as basic, but due to ESS it is easy to
recalculate them for basic SF of any order. We also plotted $\alpha$, 
the weakening factor exponent, calculated using SFs of different
order. We see that $\alpha$ is generally smaller than 1/4.

%\placefigure{interm}
\begin{figure}
\figurenum{3}
\plotone{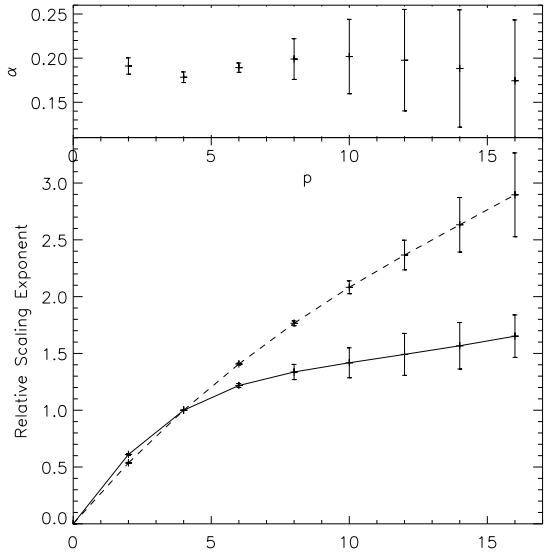}
  \caption{ Incompressible cubes, $M_A=0.7$, {\it Lower panel}: Relative scaling
exponents for SFs $<|\delta w \delta u|^{p/2}>$ (solid) and 
$<|\delta w \delta u \sin\theta|^{p/2}>$ (dashed) with $p=4$ taken as basic SF.
{\it Upper panel}: interaction weakening scaling factor $\alpha$ (see sec. 5)
calculated using different order structure functions.}
\label{interm}
\end{figure}

\section{Discussion}
In our incompressible simulations magnetic field energy was
typically slightly larger that kinetic energy (see Fig 1.)
This difference is the so-called residual energy (see Muller,
Grappin, 2005). Residual energy creates a polarization
alignment, but of different type that we measure. In this case
${\bf w}$ and ${\bf u}$ are systematically {\it antiparallel}. However,
we are looking not for difference between p.d.f. of angles of 0
and 180 degrees, but for difference for 0 and 90 degrees. It would be
a second-order correction and is generally pretty small, as long
as kinetic and magnetic energies are close. We estimated that in our
case this spurious effect is never larger than 2\%.

Maron and Goldreich (2001) observed axial asymmetry or net polarization of
Alfv\'en waves. They claimed that this effect is strongest in decaying
turbulence, but in forced turbulence it is bound. They did not, however,
connected this effect to the flatter than $-5/3$ spectra observed in
both decaying and forced runs. They speculated that spectra could be
flatter due to intermittency.

Boldyrev (2005) considers weakening of interaction that comes from
three-dimensional structure of the eddy. As the model
he uses same Kolmogorov-type arguments, so his formulae
can be reproduced by ours by taking $\alpha=\alpha'/(3+\alpha')$,
where $\alpha'$ is Boldyrev's $\alpha$.
However, in our simulations we saw that purely geometrical
arguments can not fully describe weakening
and the real weakening is due to correlation between angular
alignment and the wave strength.

Somewhat unexpected result is that compressible cubes show
more of a polarization alignment, even though they naturally have
a smaller inertial range due to large numerical diffusivity and viscosity.
So far within our approach $\alpha$ is
not constrained and we do not claim that weakening will nessesarily
behave as a power-law. 
We leave this two issues to further study.

While the magnitude of the changes of the spectral index as well as anisotropy
that we observe is not large, the potential implications of this can be very 
substantial. Turbulence in interstellar medium is injected on the scale
of dozens of parsecs (see Lazarian \& Pogosyan 2000, Farmer \& Goldreich 2004),
and on the megaparsec scales in clusters of galaxies (see Cassano \& Brunetti
2005). The scale at which Alfven turbulence dissipates can be of the order
of thermal proton Larmor radius, i.e. thousands of kilometers. As the result
of such a humongous scale separation any changes in the spectra and 
scale-dependent anisotropy will have
important consequencies for the turbulent energy available at a sufficiently
small scale. Cosmic ray acceleration and propagation are the processes
that are directly affected.

\section{Summary}
In the paper above we have demonstrated that

1. Turbulent magnetized flow spontaneously develop regions where the
mutual shearing of the oppositely moving Alfv\'en waves is weakened
due to polarization alignment of the waves.

2. The amplitude of waves is enhanced in the regions of correlated polarization.
    
3. Even though the effect of polarization alighnment is weak, it affects
spectrum and anisotropy, which could have a significant impact on a wide
range of astrophysical phenomena.

{\acknowledgments
AL thanks Ethan Vishniac fot stimulating discussion of polarization
effects during 2002 Ringberg workshop.
AB thanks IceCube project for support of his research.
AL acknowledges the  NSF grant AST-0307869 and the support from
the Center for Magnetic Self-Organization in Laboratory and Astrophysical
Plasma.
}


\begin{thebibliography}{}
\bibitem{Arm}
Armstrong, J.W., Rickett, B.J., Spangler, S.R. 1995, ApJ, 443, 209

\bibitem{Benz}
Benzi, R. et al 1993, Phys. Rev. E 48, R29

\bibitem{Bisk}
Biskamp, D. 2003, {\it Magnetohydrodynamic Turbulence.}\, (Cambridge: CUP)

\bibitem{Bold}
Boldyrev, S. 2005, ApJ, 626, L37

\bibitem{Cass}
Cassano, R., Brunetti, G. 2005, MNRAS, 357, 1313

\bibitem{Chandr}
Chandran, B. 2000, Phys. Rev. Lett., 85, 4656

\bibitem{CLV02} 
Cho, J., Lazarian, A. 2002, Phys. Rev. Lett. 88, 24, 245001 

\bibitem{CL03}
Cho, J., Lazarian, A. 2003, MNRAS, 345, 325

\bibitem{CL03}
Cho, J., Lazarian, A. 2006, ApJ, in press

\bibitem{CLV02} 
Cho, J., Lazarian, A., \& Vishniac, E. 2002, ApJ, 566, L49 

\bibitem{ChoV00}
Cho, J. \& Vishniac, E. 2000, ApJ, 539, 273

\bibitem{Farm}
Farmer, A.J., Goldreich, P., 2004, ApJ, 604, 671

\bibitem{Galt}
Galtier, S., Nazarenko, S.V., Newell, A.C., Pouquet, A. 2002, ApJ, 564, L49

\bibitem{GolS95}
Goldreich, P., \& Sridhar, S. 1995, ApJ, 438, 763 (GS95)

\bibitem{Hig}
Higdon, J.C. 1984, ApJ, 285, 109

\bibitem{Horb} Horbury, T.S. 1999, in Plasma Turbulence and Energetic Particles,
eds. M. Ostrowski and R. Schlickeiser, Cracow, p. 28

\bibitem{LP00}
Lazarian, A., Pogosyan, D. 2000, ApJ, 537, 720

\bibitem{LitG01}
Lithwick, Y., \& Goldreich, P. 2001, ApJ, 562, 279

\bibitem{Mar}
Maron, J., Goldreich, P. 2001, ApJ, 554, 1175

\bibitem{MATT}
Matthaeus, W.H., Brown, M.R. 1988, Phys. of Fluids, 31, 3634

\bibitem{Mon}
Monin, A.S., Yaglom, A.M. 1975, {\it Statistical Fluid Mechanics, vol. 2}
%: Mechanics of Turbulence
(MIT Press, Cambridge, MA)

\bibitem{Mull}
Muller W.-C., Grappin, R 2005, Phys. Rev. Lett., 95, 114502

%\bibitem{Kol41}
%Kolmogorov, A. 1941, Dokl.~Akad.~Nauk SSSR, 31, 538

\bibitem{Schl}
Schlickeiser, R. 2002, {\it Cosmic ray astrophysics.}\, (Berlin: Springer)

\bibitem{Sheb}
Shebalin, J.V., Matthaeus, W.H., Montgomery, D. 1983, J. of Plasma Phys., 29, 525

\bibitem{YL02}
Yan, H., Lazarian, A. 2002, Phys. Rev. Lett., 89, 281102

\bibitem{YL04}
Yan, H., Lazarian, A. 2004, ApJ, 614, 757

\end{thebibliography}
\end{document}